\documentstyle[twocolumn,aps]{revtex}
\begin{document}
\draft
\twocolumn[\hsize\textwidth\columnwidth\hsize\csname @twocolumnfalse\endcsname
%
%
%

\title{Ferromagnetism in Electronic Models for 
Manganites}

\author{ Jose Riera$^1$, Karen Hallberg$^2$, and Elbio Dagotto$^3$}

\address{
$^1$ Instituto de F\'{i}sica Rosario y Depto. de F\'{i}sica, Univ. Nac. de
Rosario, 2000 Rosario, Argentina
\\
$^2$ Max-Planck-Institut f\"ur Physik Komplexer Systeme, Bayreuther
Strasse 40, D-01187 Dresden, Germany
\\
$^3$Department of Physics and National High Magnetic Field Lab,
Florida State University, Tallahassee, FL 32306, USA}

\date{\today}
\maketitle

\begin{abstract}
Ground state properties of the Kondo model for manganese
oxides in one dimension are studied using numerical techniques. 
The large Hund coupling ($J_{H}$) limit
is specially analyzed. A robust region of fully saturated ferromagnetism (FM)
is identified at all densities. For open boundary conditions it is shown
exactly that the ground state is FM at $J_{H} = \infty$.
Hole-spin 
phase separation  competing with
FM was also observed when a large exchange $J$ between the ${\rm Mn^{3+}}$
ions is used. As the spin of the
transition metal ion grows, the hole mobility decreases providing a
tentative explanation for the differences between Cu-oxides and Mn-oxides.

\end{abstract}

\vskip2pc]
\narrowtext

%
%

Doping of transition metal oxides with perovskite structure
induces remarkable phenomena such as high temperature
superconductivity, charge ordering, and anomalous transport
properties. Typical examples are the layered
Cu-oxides and Ni-oxides. Recently, another dramatic property
of doped carriers in transition metal oxides has been revealed: 
at low temperature ($T$) ${ La_{1-x} Ca_x Mn O_3}$  changes from
an antiferromagnetic (AF) insulator to a ferromagnetic metal
as $x$ grows from 0. 
``Colossal'' magnetoresistance phenomena
is observed as $T$ increases from the ferromagnetic phase\cite{jin}.
At the critical ferromagnetic temperature, $T^{FM}_c$, 
a metal-insulator transition occurs
and the high temperature regime is a paramagnetic  insulator.
At larger doping $x>0.5$, 
a charge-ordered AF state was detected\cite{phase}.

The double-exchange (DE) mechanism has been used to
explain the FM phase in Mn-oxides\cite{zener},
since the mobility 
of the $e_g$ electrons increases for a polarized background 
of localized $t_{2g}$ electrons.
However, recent reexamination
of the DE model, where electrons move with an effective hopping
proportional to the cosine of half the angle between the 
localized spins (assumed classical),
 revealed that electrons can actually
collect a Berry phase that may induce states of low energy in the
spectrum\cite{muller}.
This  effect could  reduce the  $T^{FM}_c$ and alter the
behavior of the system at higher temperatures. 
In addition, from recent numerical studies of the Kondo Hamiltonian
for Mn-oxides it was concluded that the ground state at nonzero hole density
is not FM but a spin singlet, although with
local ferromagnetic correlations\cite{bishop}. These results suggest that 
electronic models for the Mn-oxides more realistic than the DE model
may have a phase diagram richer than expected which deserves 
to be studied with state-of-the-art
computational techniques. Such studies would clarify, among other issues, 
whether FM indeed exists
in the $T=0$ ground state and what phases are in competition with it.
In addition, it would be important to analyze the origin of  the
drastic differences between Cu-oxides and Mn-oxides regarding their
doped-induced properties. The cuprates
have metallic and superconducting phases,
while the manganites are either
charge-ordered insulators or ferromagnets\cite{phase}.
In this paper, both
issues are addressed.

The Hamiltonian widely used for Mn-oxides is\cite{zener}
$$
{H = -t \sum_{\langle {\bf m}{\bf n} \rangle \sigma} (c^\dagger_{{\bf m}
\sigma} c^{\phantom{\dagger}}_{{\bf n} \sigma} + H.c.) - J_{H} 
\sum_{\bf n} {{\bf\sigma}_{\bf n}}\cdot{{\bf S}_{\bf n}}. }
\eqno(1)
$$
\noindent Here the first term represents the $e_g$-electron 
transfer between nearest-neighbor Mn-ions at sites ${\bf m}$ and ${\bf n}$.
The second
term corresponds to the Hund coupling ($J_{H} > 0$) 
between the $S=3/2$ $t_{2g}$ localized spin
${\bf S_{n}}$ and
the spin ${\bf \sigma_{n}}$ of the mobile $e_g$-electron at the same site.
Coulombic repulsion in the $e_g$-band is not included in Eq.(1) but it is
considered in some calculations below\cite{phonons}. 
Phenomenologically, it is expected that
$J_{H} \gg t$ which favors the alignment
of the itinerant and localized spins. For ${\rm Mn^{3+}}$, where the
$e_g$ level is occupied, the resulting 
spin is 2, while for ${\rm Mn^{4+}}$ (vacant $e_g$ state) the spin is 3/2.
Thus, at $J_{H} = \infty$ the effective 
degrees of freedom become $S=2$ ``spins'' and 
$S=3/2$ ``holes'' which is  the language convention  followed below. 
In this limit Eq.(1) 
reduces to a simpler hole-hopping Hamiltonian. For ``spin'' values $S \ge 1$,
the hopping term at
$J_{H} = \infty$ is\cite{muller}
$$
{  H_{J_{H}=\infty} = -t \sum_{ \langle {\bf m}{\bf n} \rangle} P_{{\bf m}{\bf n}}
~Q_{S}(y),
}
\eqno(2)
$$
where $y= {{{\bf S}_{\bf m}}\cdot{{\bf S}_{\bf
n}}}/ S (S-1/2)$ and for Mn-oxides ($S=2$) the polynomial $Q_S (y)$ is
$$
{ Q_{2}(y) = -1 - {{5}\over{4}} y + {{7}\over{4}} y^2 + {{3}\over{2}} y^3 .
}
\eqno(3)
$$
For Ni-oxides ($S=1$) $Q_1 (y) = (1+y)/2$\cite{nio}.
Hamiltonian Eq.(2) is rotational invariant and
it acts only on links containing one spin and one hole. 
${ P_{\bf mn} }$  permutes the hole and the spin.
The ${\bf S}_{\bf n}$ operators are standard and in Eqs.(2,3)  they
act over both
the spin and hole involved in the hopping process, i.e. the kinetic term
 not only
interchange their site positions but 
it can flip their spin projections as well\cite{muller,nio}.
This is an important difference with respect to the standard $t-J$ model.
At finite but large $J_{H}$, Eq.(2) is supplemented by a
Heisenberg interaction between nearest neighbor spins of the form
$J \sum_{\langle {\bf mn} \rangle} ({{\bf S}_{\bf m}}\cdot{{\bf S}_{\bf n}} 
- n_{\bf m} n_{\bf n}/4)$, in the standard notation.\cite{review}
This term only affects the S spins since in hole states
the orbital   with mobile electrons ($e_g$) is empty.

As numerical techniques the exact diagonalization\cite{review} (ED) 
and the infinite and finite-size 
density matrix renormalization group\cite{white} 
(DMRG) methods were used.
The special case $S=1$ will be studied 
together with the relevant case for Mn-oxides ($S=2$) to observe the
effect of different transition metal spins in the phase diagrams.
In addition, experience with Cu-oxides models shows
that the analysis of Hamiltonians in 1D chains provides
insight on qualitative features of the phase diagram that
survive the increase in
dimensionality. 
Since 1D is the best for numerical techniques, 
our study below is restricted to chains\cite{comm4}.
Note that the special case
$S=1$ and 1D has intrinsic relevance since the 
compound ${ Y_{2-x} Ca_x Ba Ni O_5}$
with NiO chains 
has been recently synthesized\cite{batlogg}.


First let us show
that the ground state of Eqs.(1,2) in 1D is
a fully saturated ferromagnet (FM) 
for hole density $x \neq 0,1$ in the case of $open$
boundary conditions (OBC) and $J_{H}=\infty$ (or $J=0$). 
The proof is a simple extension of Kubo's results for the DE
model\cite{kubo}: 
the matrix elements
of the $J=0$ Hamiltonian Eq.(2) in the standard $S^z$-basis
are $nonpositive$, which is shown by explicitly writing 
the polynomials $Q_s(y)$ in matrix form for a given link. 
For a subspace with a fixed total
spin projection $S^z_{total}$, this property implies that the ground state is
nondegenerate and the coefficients of this state expanded in the
$S^z$-basis are of the same sign(Perron-Frobenius theorem)\cite{auerbach}.
Since in each subspace with a fixed
$S^z_{total}$ the
state $| S_{max}, S^z_{total} \rangle $ 
with the maximum possible spin $S_{max}$
has all of the coefficients of the same sign, 
then the ground state in such subspace
must be the ferromagnetic state. Q.E.D.
As in Kubo's results, this proof is not valid for $twisted$ boundary
conditions, including periodic and antiperiodic conditions (PBC and APBC, 
respectively),
since in this case across the boundary it is possible to move
fermions from site N to 1 (for a N-site chain)
collecting anticommutation signs which make the matrix elements
not necessarily negative. For the same reason
the proof is not extended to higher dimensions even with
OBC. Another important detail is that each $S^z_{total}$ subspace
is assumed to be fully connected (i.e. all
elements of the basis are reached by successive applications of the
Hamiltonian over any of its members). Actually in the
1D $t-J$ model, the
Hilbert space is disconnected\cite{nagaoka} 
for $J=0$ and the theorem does not hold.
However, if $S \ge 1$ the Hilbert space is
$connected$ since together with a hopping, the
spins and holes can flip
their spin projections\cite{ejemplo}.

Solving exactly Eq.(2) on finite chains with OBC we 
verified the existence of ground state FM
for any $x \neq 0,1$. 
We also observed that using PBC(APBC) for an odd(even)
number of holes (closed shell) for 
both $S=1$ and $2$, the ground state at $J=0$ is
again a FM. 
Thus, to increase the size of the chains that can be studied
numerically with ED techniques, preserving the correct properties of
the ground state, below closed-shell BC have been used 
in addition to OBC. On the other
hand, using PBC for an even number of holes, the ground state is
a spin singlet\cite{bishop}.
Thus, in principle these BC should not be used for finite cluster
studies of $T=0$ properties\cite{comm2}. 
However, the boundary dependence of the
ground state total spin on finite chains 
suggests the existence of
low energy singlet excitations  close to the ground 
state\cite{muller,bishop}.

In Fig.1a the phase diagram for $S=1$ is shown.
Chains with up to 16 sites were studied with ED.
Using the closed-shell BC, a robust region of FM
was identified.
The FM boundary was also calculated 
using the DMRG technique
at $x=0.25$, $0.50$ and  $0.75$ with chains of up 
to 40 sites and OBC\cite{dmrg}. The
critical coupling to reach FM obtained
with ED and DMRG are in excellent agreement, showing that size effects
are small\cite{partial}.

Also in Fig.1a, regions of phase separation (PS) and hole binding (B)
have been identified using ED 
by calculating the compressibility and binding energy from
$E(M \pm 2)$, $E(M \pm 1)$, and
$E(M)$, where $E(M)$ is the ground state energy 
for $M$ holes\cite{comm1}. 
PS is characteristic of spin-hole models, although not contained
in the DE model, and it occurs
at large $J/t$ when the kinetic energy gained by making holes mobile
is overcome by the  magnetic energy lost. This has been observed in
the  $t-J$ model in both 1D and 2D\cite{review} 
and it is not surprising to
find a similar behavior in models with a larger $S$. However, 
at $x=0.5$ PS appears at $J \sim 1.1$ which is much smaller  than the coupling
$J/t \sim 3.0$ needed
in the $t-J$ model\cite{ogata}, suggesting that electronic 
PS may play an important
role in the physics of Mn-oxides. 
Close to PS there is a region (B)
where holes form mobile pairs.
Superconductivity is likely in
this regime which also appears in the $t-J$ model\cite{haldane}.

In Fig.1b, the $S=2$ phase diagram is shown. Here the
chains accessible to ED have 8 sites, while with
DMRG we studied up to 30 sites. 
The finite-size effects on the FM line are small.
As for $S=1$, a robust regime of FM has
been identified. We observed that the FM lines of phase transitions 
of Fig.1a,b  are very similar
if $J$ is scaled as $JS(S+1)$. 
PS and B are also present. Note that for $S=2$ the intermediate
metallic regime is very narrow, i.e. as $S$ grows the trend observed in Fig.1a,b
suggests
that the PS and FM phases become dominant. To understand this effect note
that the ``vaporization'' of a large spin cluster in the PS region,  
held together by 
magnetic forces of effective strength $JS(S+1)$, 
can only occur if a
substantial gain in kinetic energy, dominated by the scale $t$, 
is achieved by the vaporized 
spins in the hole rich region.
Thus, the critical $J/t|_{PS}$ 
leading to PS should follow $J/t|_{PS} \sim 1/S(S+1)$, or a faster decay with
$S$ once the reduction in hole 
kinetic energy due to the scattering with the magnetic background is
considered. This argument breaks down at very
small $J/t$ where the FM state is energetically competitive.
Thus, in the large $S$
limit the metallic regime at $T=0$ would disappear in favor of PS and FM
phases.

The study of the intermediate metallic
regime is interesting
since it may contain features of the paramagnetic insulator
observed in manganite experiments at $T > T^{FM}_{c}$.
As $S$ grows, a tendency towards localization is indeed
observed in this regime.
For example, in Fig.2a the bandwidth $W$ of $one$ hole for
$S=1/2, 1,$ and $2$ vs $t/JS(S+1)$ is shown. It is clear that $W$ rapidly
diminishes with $S$.
Similar effects are observed in Fig.2b where the kinetic energy in the
metallic region is presented at $x=0.5$.
Such a mobility reduction is
reasonable since holes destroy the magnetic order in their
movement, paying an energy that rapidly grows with the spin. The
large effective mass
for $S=2$ indicates that small perturbations away from
a translational invariant system may localize the carriers.
This effect and the dominance of PS and FM at large S
is a possible explanation of the notorious experimental
differences at low-T between Cu-oxides, which show metallic 
and superconducting 
phases, and
Mn-oxides, which have FM and charge-ordered phases\cite{phase}
(note that previous studies in Ni-oxides
have shown that charge-ordering and phase separation are
related phenomena\cite{previous}).

The Hamiltonian Eq.(1) does not include a spin-spin
exchange between the $t_{2g}$ spins since 
such a coupling arises indirectly from Eq.(1) mediated by the $e_g$ electrons.
However,  as $x \rightarrow 1$ this RKKY
coupling vanishes, which is not in agreement with experiments that show
AF order at $x=1$. To remedy this problem phenomenologically, a Heisenberg
interaction with coupling $J'$
between the ``holes'' (which carry spin) 
can be introduced. At $x=1$ now
the ground state would be AF as for $x=0$. For completeness, the numerical
analysis of Fig.1a was repeated adding a term
$J' \sum_{\langle {\bf mn} \rangle} ( {\bf
S}_{\bf m} \cdot {\bf S}_{\bf n} - n_{\bf m} n_{\bf n}/4)$ 
between the $S=1/2$ holes and considering $J'=J$ (see
Fig.3a). Up to $x \approx 0.5$ the phase diagram is similar to the
results of Fig.1a, but as $x$ grows further 
the coupling needed to reach the FM phase is reduced, as expected. 
FM remains robust at all densities and small $J/t$.

Finally,
in Fig.3b the phase diagram of Hamiltonian Eq.(1) solved
exactly on a 4-site chain and using DMRG on
an 8-site chain, both with OBC and at $x=0.5$, 
is shown. Here, an on-site
Hubbard repulsion $U/t$ for $e_g$ electrons is
incorporated in the Hamiltonian. In agreement with our analysis of the
$J_{H} = \infty$ limit, the ground state is FM at
large $J_{H}$. As $U/t$ grows, FM becomes stable even at small
$J_{H}$ which is natural
since in the Hubbard model ferromagnetism
tends to be favored in strong coupling. As with model Eq.(2),
the non-FM region contains a small ferromagnetic component, i.e. for
$J_H = U =1$ and $N=8$ using DMRG we find 
spin one in the ground state.

Summarizing, a numerical study of the 1D Kondo Hamiltonian for Mn-oxides
 in the  $J_{H} \gg t$ limit and also at finite $J_{H}$ 
has been presented.
The phase diagram contains a robust FM
 region at all densities
in agreement with expectations from the DE model, and
contrary to the case of the 1D $t-J$ model for Cu-oxides.
For OBC we showed exactly that the ground state 
at $J_H = \infty$ and $x \neq 0,1$ is FM  with maximum spin.
At finite exchange $J$, 
a narrow metallic window with low mobility carriers
separates the FM region from a phase separated
regime, not contained in the DE model.
As the spin S grows, PS and FM become dominant, the intermediate
metallic regime tends to disappear, and
the mobility of holes  in the non-FM region
rapidly decreases.

We thank E. M\"uller-Hartmann, K. Penc, 
A. Sandvik, and P. Thalmeier for useful conversations.
E. D. is supported by the 
NSF grant DMR-9520776.

\medskip

\vfil

%
%

{\bf Figure Captions}

\begin{enumerate}

\item (a) Phase diagram of the $S=1$ Hamiltonian Eq.(2) 
on a 1D chain ($J S (S+1)$ is
the coupling scaled with the spin, $x$ the hole density, and $t=1$). 
The gray line and open points denote the FM phase boundary.
Open circles(10), rotated triangles(10), 
squares(12), diamonds(14), and triangles(16) 
correspond to ED results 
using closed-shell BC (in parenthesis the number of sites).  
DMRG results for the FM boundary with OBC
and $N=40$ are also shown (stars).
The solid line joining full circles is
the PS boundary calculated with ED ($N=12$).
A hole binding (B) region exists between
the dashed line and PS; (b)
Same as (a) but for the $S=2$ Hamiltonian. 
The boundaries  of PS and FM were 
evaluated with ED on chains of $N=8$ sites.
DMRG results on $N=30$ OBC are shown (stars).
In the intermediate metallic region there is binding.

\item
(a) Bandwidth $W/J$ vs $t/JS(S+1)$
for $one$ hole and $S=1/2$ (full diamonds), $1$ (open circles), 
and $2$ (full squares) calculated using ED techniques
on $N=12, 12,$ and $8$ chains, respectively; (b)
Kinetic energy (hopping term ground state expectation value) 
in units of $t$ vs $t/JS(S+1)$ obtained 
at $x=0.5$ for $S=1/2, 1,$ and $2$ using $N=12,12,$ and $8$ chains,
respectively.

\item (a) Same as Fig.1a but adding 
a Heisenberg interaction between the $S=1/2$ holes
of strength $J'=J$; (b) Phase diagram
of the Kondo model Eq.(1) at $x=0.5$
including an $e_g$ Hubbard repulsion of strength $U$ ($t=1$).
Open (full) squares are results obtained with ED (DMRG) on chains with
N=4 (8) sites. The non-FM region indicates weak ferromagnetism in the
ground state, according to the DMRG results.

\end{enumerate}


\begin{references}
\bibitem{phonons} Phononic effects, which 
may be important for the manganites, are not 
incorporated in the present analysis.
Also the effect of orbital degeneracy
has been neglected
(W. Koshibae, Y. Kawamura, S. Ishihara, S. Okamoto, J. Inoue, and S.
Maekawa, preprint).
\bibitem{ejemplo}
For example, consider a 3-site OBC problem
for $S=1$ and an initial state 
$ | 1/2,-1/2,1 \rangle $ (with numbers representing $z$-direction spin
projections): 
here four applications
of Eq.(2) connect this state with, e.g., 
$| -1/2,1/2,1 \rangle $. On the other hand,
for the $t-J$ model the states $| 1/2,-1/2,0 \rangle $ and
$| -1/2,1/2,0 \rangle $ are disconnected.
\bibitem{dmrg} To find the boundaries of the FM phase with DMRG
we compared
the energies of different subspaces with fixed 
$S^z_{total}$.
The FM case is very precise, having a truncation error of $10^{-12}$
for $N=100$
when keeping m=200 states per block. The states with smaller $S^z_{total}$
are less precise due to a larger Hilbert space. For $S=1$ ($S=2$) we used 
m=200 (m=100), N=40 (30), with a truncation error $10^{-6}$ ($10^{-4}$).
\bibitem{partial} In the metallic phase
limiting  the FM phase, $weak$ ferromagnetism has been
observed using the  DMRG method. For example,
at $J/t=1,x=0.5$, the state with
$S_{total}^z =1$ is degenerate with $S_{total}^z =0$ for clusters with
$N \ge 12$ and OBC. However, the spin per site is so small that further
work is needed to verify that this effect survives the bulk limit.
\bibitem{comm1} 
Using DMRG we could not obtain with enough accuracy the boundary of
the PS region for larger systems
 due to large uncertainties in the data. For the
treatment of ground states with spatial charge ordering it is
necessary to modify the standard DMRG method.
\bibitem{haldane} In addition to PS, B, and FM, close to $x=0$ in 1D 
there must
be a region with a spin-gap for both $S=1$ and $2$, which was not
explored here.
\end{references}
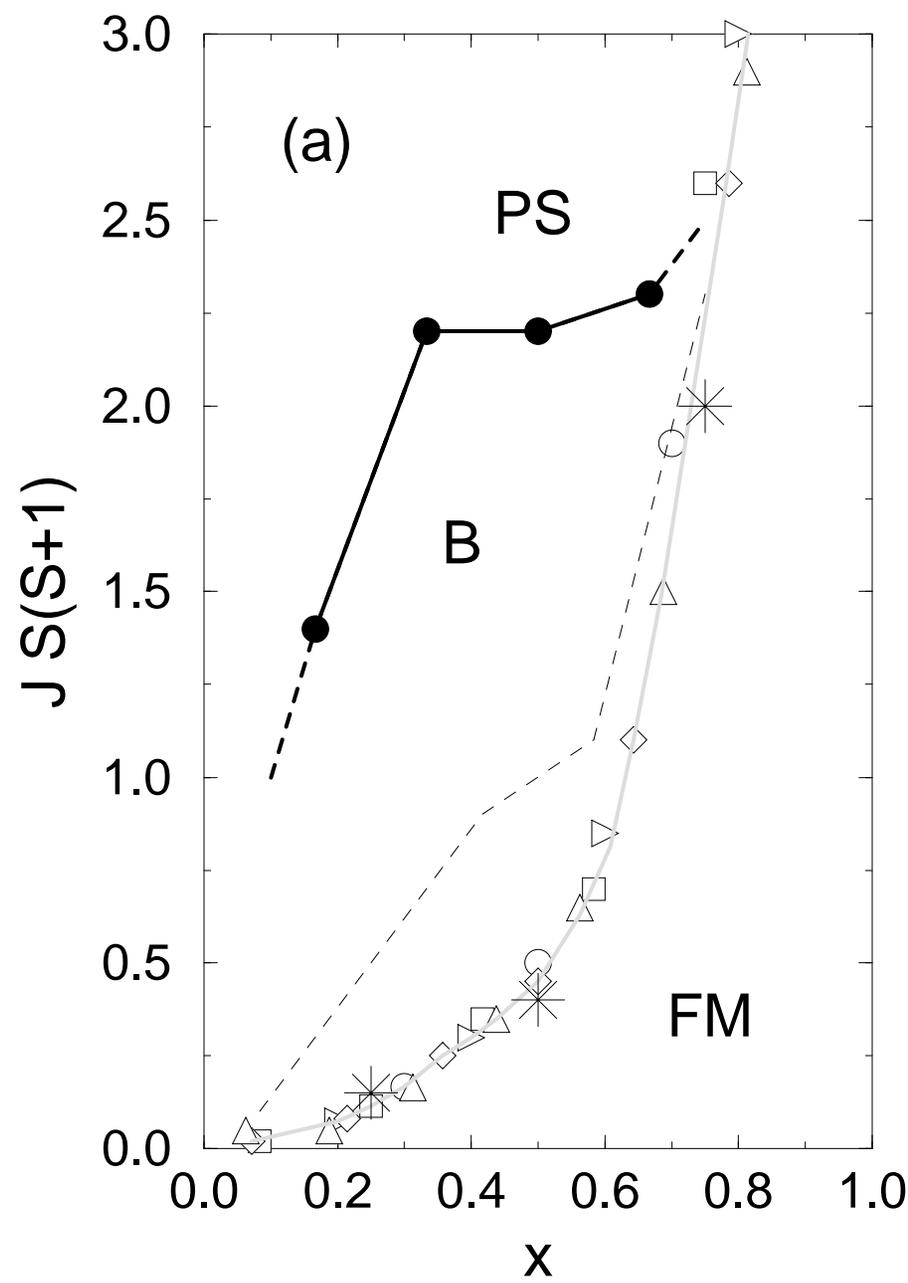
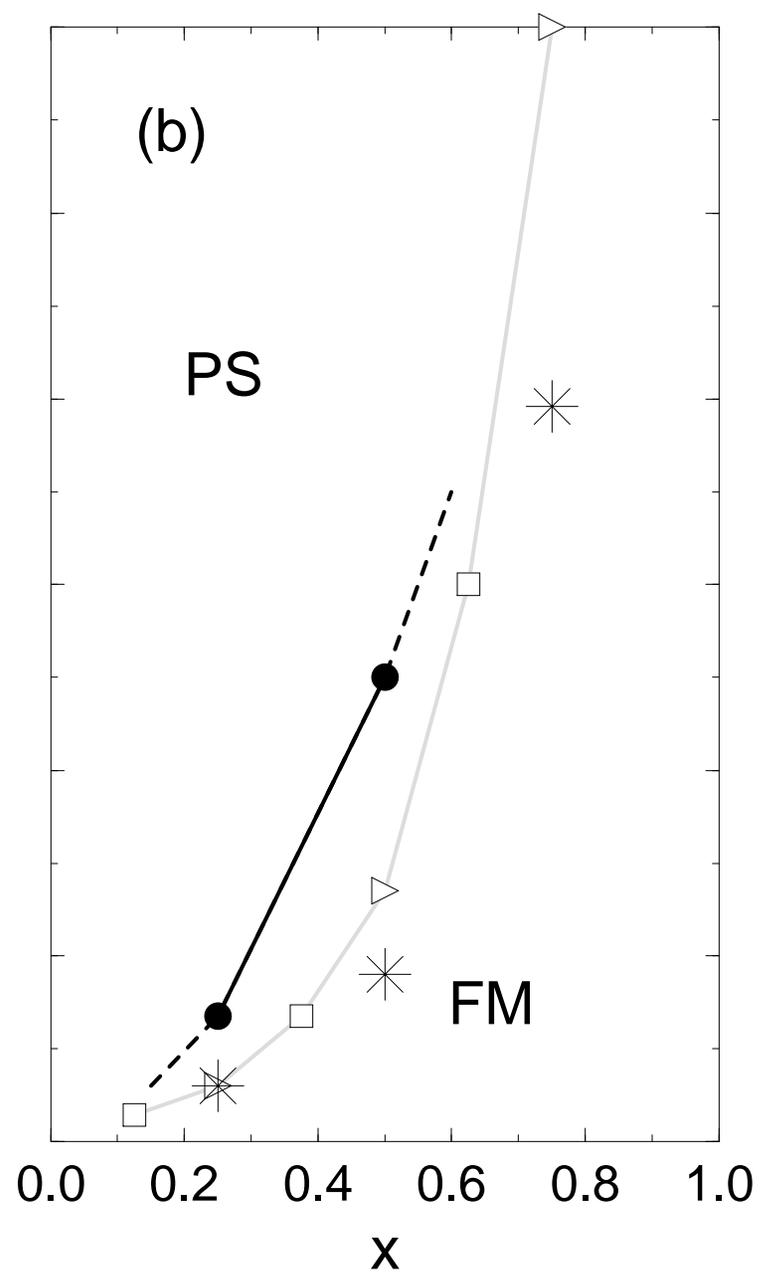
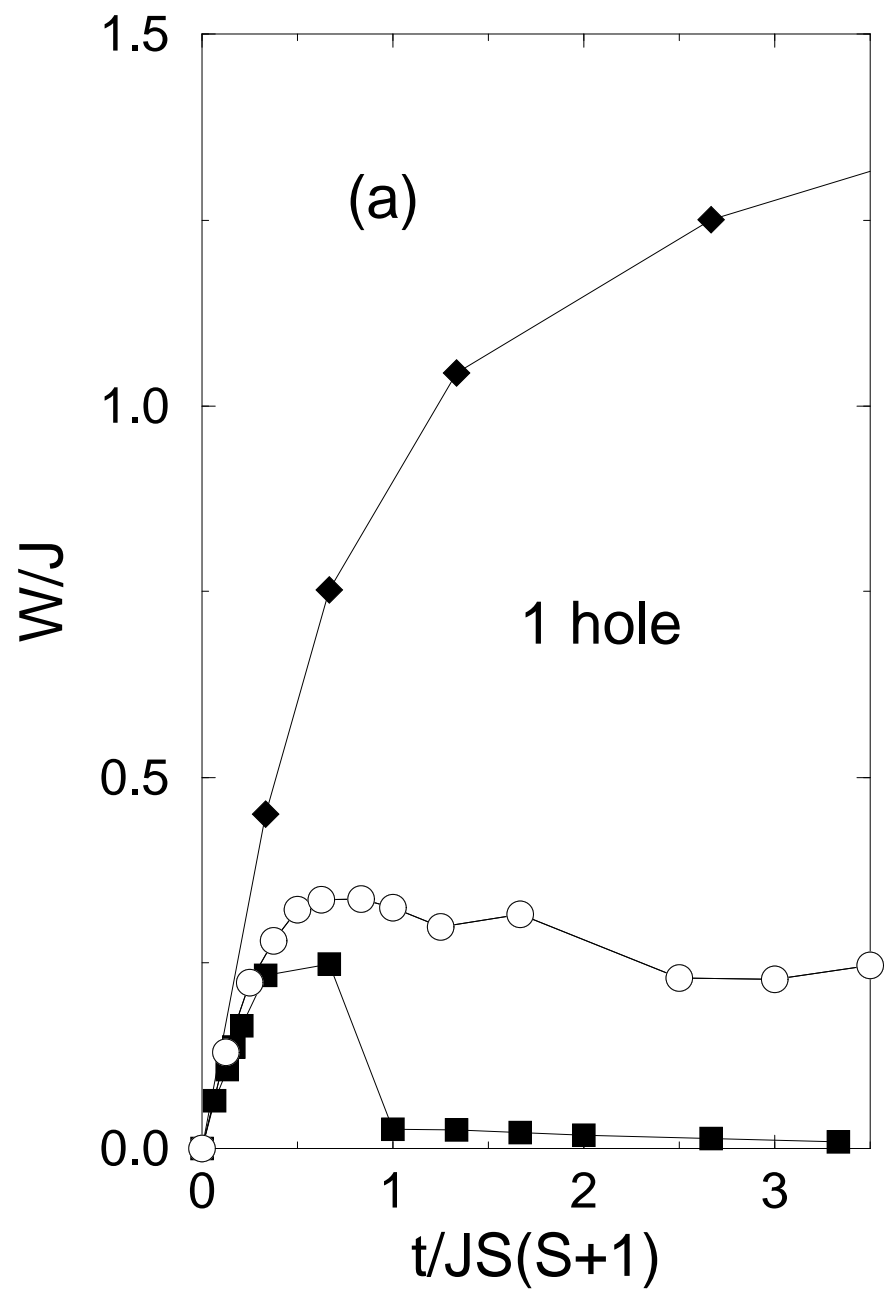
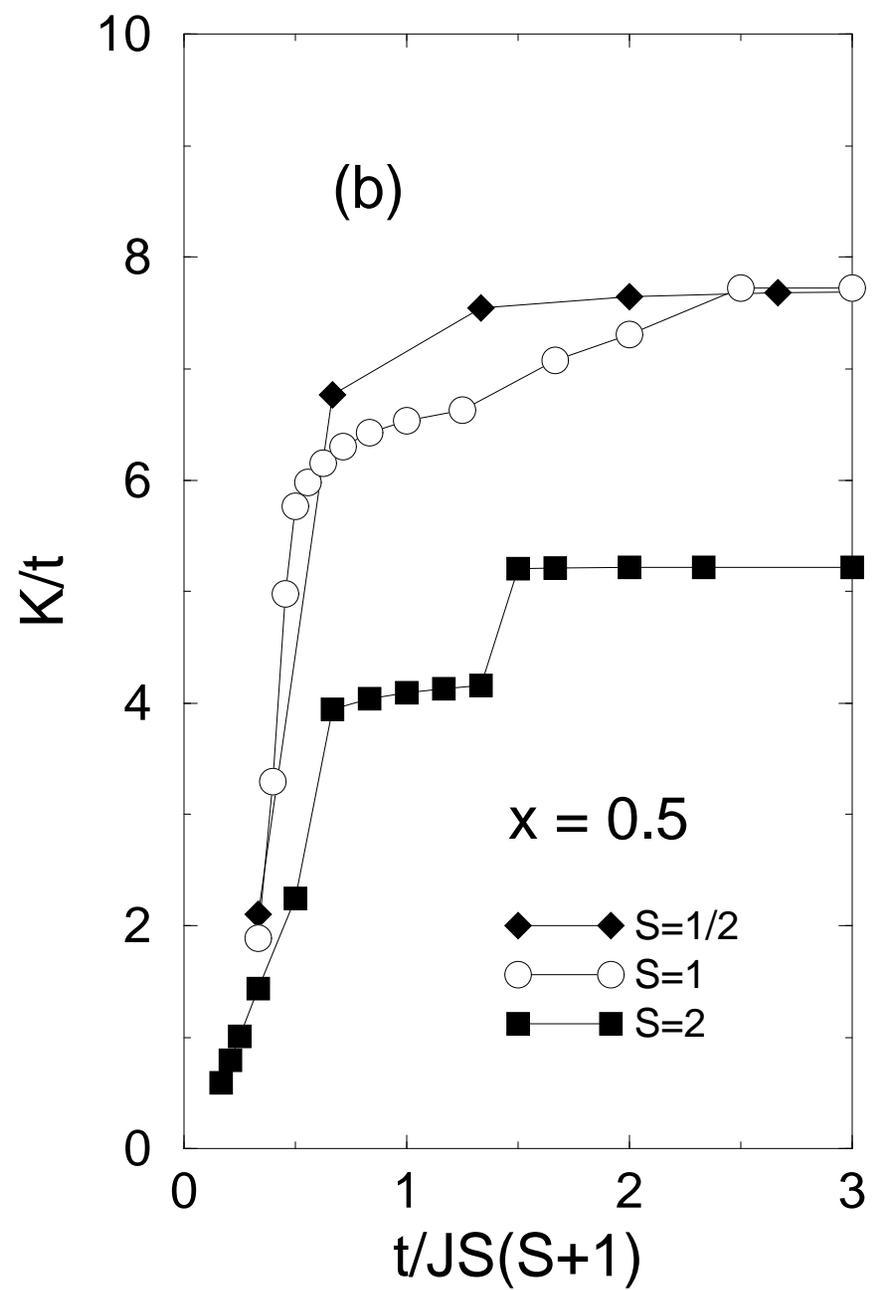
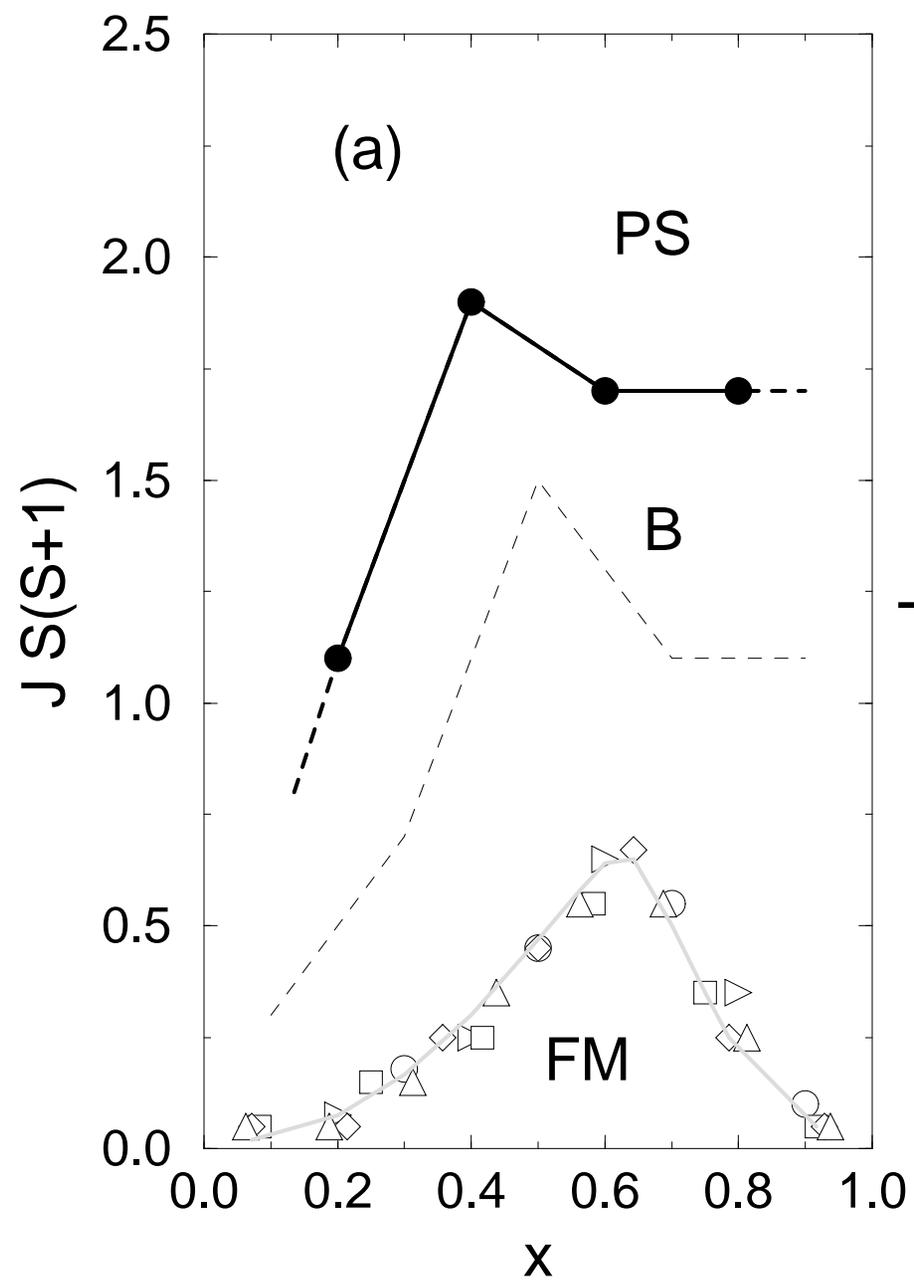
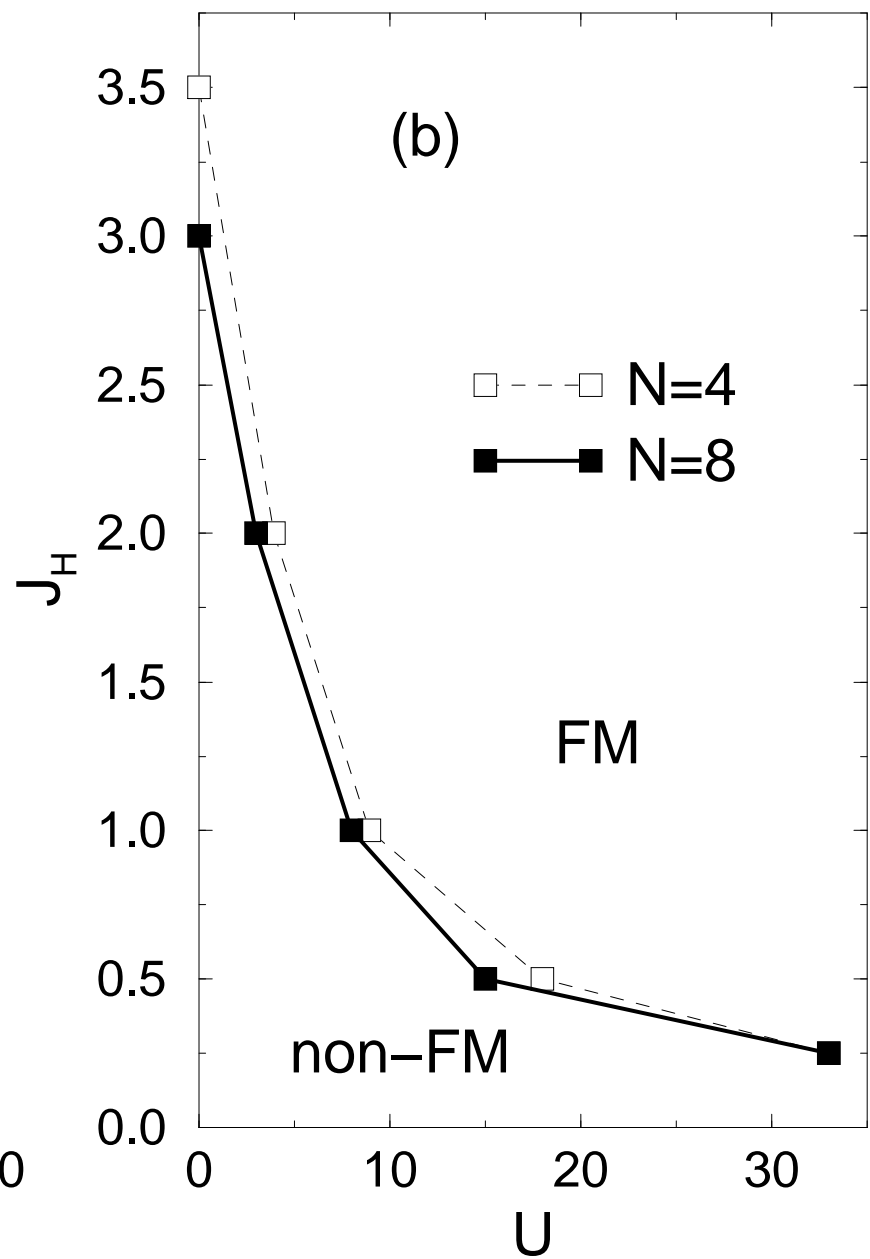
\end{document}